# Electromagnetic Hydrophone with Tomographic System for AbsoluteVelocity Field Mapping

Pol Grasland-Mongrain[1)], Jean-Martial Mari, Bruno Gilles, Jean-Yves Chapelon, Cyril Lafon

Inserm, U1032, LabTau, Lyon, F-69003, France ; Université de Lyon, Lyon, F-69003, France

**The velocity and pressure of an ultrasonic wave can be measured by an electromagnetic hydrophone made of a thin wire and a magnet. The ultrasonic wave vibrates the wire inside a magnetic field, inducing an electrical current. Previous articles reported poor spatial resolution of comparable hydrophones along the axis of thewire. In this study, submillimetric spatial resolution has been achieved by using a tomographic method. Moreover, a physical model is presented forobtainingabsolute measurements. A pressure differential of 8% has been found between piezoelectric and electromagnetic hydrophone measurements. These characteristics show this technique as an alternative to standard hydrophones.**

Due to the growing use of ultrasonic devices, there is a need for preciseultrasound pressure mapping. During the last four decades, different types of pressure mapping devices have been developed, the most used today being piezoelectric and piezoceramic hydrophones, interferometers[1, 2], fiber-optic hydrophones[3] and Schlieren imaging systems[4].On the other hand, the Lorentz force has been applied in acoustic since 1931 with the ribbon microphone for audible sound [5]. In this setup, the sound makes a conductive ribbon to vibrate between the poles of a magnet, inducing an electrical current due tothe Lorentz force. It has also been applied to ultrasound in water in 1969 [6], using a thin wire wound around an insulator. The ultrasound wave vibrates the insulator and the wire, which gives rise to an induced electrical current.In the context of pressure field measurement, the main disadvantage of this approach is the resolution along the wire. A compromise of 5 mm has been proposed by J. Etienne et al. [7] which can still be too wide for high resolution ultrasound measurements. Y. Sharf et al. [8] improved this device by transmitting the produced electrical signal to a pick-up coil by electromagnetic coupling, but with lower sensitivity.

This article introduces a technique using the Lorentz force to quantitatively map velocity and pressure fields with a spatial resolutionof less than one millimeter. Good spatial resolution is achieved by moving and rotating the wire then making a tomographic reconstruction of the pressure field.

The movement of a conductor inside a magnetic field induces an electrical current perpendicular to the motion and to the magnetic field. For a conductor of length $L$, the induced voltage $e$ is proportional to the velocity $v_z$ of the conductor times the magnetic field $B_x$ integrated along the conductor, as given by Eq. 1

---

1) Corresponding author: pol.grasland-mongrain@inserm.fr – INSERM u1032, 151 Cours Albert Thomas, 69424 Lyon Cedex FRANCE

$$e = \int_L v_z \cdot B_x \cdot dl \quad (1)$$

If the wire is insulated, or if the conductivity of the medium is small compared to the conductivity of the wire, loss of current from the wire to the surrounding medium is negligible.

The motion inducing electrical current is due to (a) the internal vibration of charges within the wire and (b) the vibration of the wire itself due to the ultrasound. For a wire with a diameter $D = 200$ µm, at an ultrasound wavelength $\lambda \approx 1.5$ mm (corresponding to a frequency of 1 MHz with the speed of sound in water about 1500 m.s$^{-1}$), we have $\lambda >> D$. The speed of sound inside the wire being about 5000 m.s$^{-1}$, the difference of pressure inside the wire is consequently small and all of the charges can be assumed to move together: in the case presented here, the phenomenon (a) is negligible and almost all observed signal is due to the contributions of phenomenon (b). The relationship between the fluid velocity $u_0$ and the wire velocity $v$ can be evaluated theoretically [9]. In the following demonstration, the wire is modeled as cylinder of infinite length. This demonstration considers a 2D plane perpendicular to the wire and all physical quantities like force and energy are taken "per unit length".

For a plane wave, the pressure $p$ and fluid velocity $u_0$ are related by Eq.2:

$$u_0 = \frac{p}{\rho c} \quad (2)$$

with $\rho$ being the density of the fluid and $c$ the speed of sound. With $p = 1.3$ MPa, $\rho = 1000$ kg.m$^{-3}$ (water density) and $c = 1470$ m.s$^{-1}$, the fluid velocity $u_0$ is approximately equal to 0.9 m.s$^{-1}$.

The Reynolds number $Re$ gives the ratio of inertial to viscous forces for a fluidic system. With $D$ diameter of the cylinder and $v$ dynamic viscosity of the medium, $Re$ is equal to:

$$Re = \frac{u_0 D}{v} \quad (3)$$

With $D = 200$ µm, $v = 1.10^{-6}$ m$^2$.s$^{-1}$ (water at 20°C) and $u_0 = 0.9$ m.s$^{-1}$ (given by Eq.2), the corresponding Reynolds number is 180. This value indicates that viscous forces are negligible compared to pressure forces, therefore the fluid can be considered as inviscid. Considering that the cylinder is small compared to the ultrasound wavelength and small turbulences through time (compatible with Re < 1000), the flow far from the cylinder is irrotational with a constant velocity $u_0$. If the typical dimensions of pressure variations (~$\lambda$) are far bigger than the dimensions of the obstacle (~$D$), the fluid can be considered as incompressible. Under these conditions, in a referential where the fluid far from the cylinder is motionless, the velocity of the fluid $\tilde{\mathbf{u}}$ is equal to the gradient of a potential $\varphi$ solution of the Laplace equation:

$$\Delta \varphi = 0 \quad (4)$$

The solutions for $\varphi$ which satisfy the boundary conditions are $\ln(r)$ derivatives, where $\mathbf{r} = r\mathbf{n}$ is the radius vector orthogonal to the cylinder axis, $R_0$ the cylinder radius and $\tilde{\mathbf{v}}$ the velocity of the cylinder in the considered referential:

$$\varphi = -\frac{R_0^2}{r^2} \tilde{\mathbf{v}} \cdot \mathbf{r} \quad (5)$$

$$\tilde{\mathbf{u}} = \nabla \varphi = \frac{R_0^2}{r^2}\left(\frac{2\mathbf{r}(\tilde{\mathbf{v}} \cdot \mathbf{r})}{r^2} - \tilde{\mathbf{v}}\right) \quad (6)$$

A more convenient relationship between $\tilde{\mathbf{u}}$ and $\tilde{\mathbf{v}}$ can be found by using the total kinetic energy of the fluid $E$, defined in Eq.7 where $\rho$ is the density of the fluid:

$$E = \frac{\rho}{2} \int \tilde{u}^2 dV \quad (7)$$

$E$ is integrated on a cylinder of large radius $R$ covering the whole space minus the wire volume (cylinder of radius $R_0$). $\tilde{u}^2$ can be substituted by $\tilde{v}^2 + \tilde{u}^2 - \tilde{v}^2$:

$$E = \frac{\rho}{2} \int \tilde{v}^2 dV + \frac{\rho}{2} \int (\tilde{\mathbf{u}} + \tilde{\mathbf{v}})(\tilde{\mathbf{u}} - \tilde{\mathbf{v}}) dV \quad (8)$$

In the second integral, $(\tilde{\mathbf{u}} + \tilde{\mathbf{v}}) = \nabla(\varphi + \tilde{\mathbf{v}} \cdot \mathbf{r})$, and by using continuity equations $\nabla \cdot \tilde{\mathbf{u}} = 0$ and $\nabla \cdot \tilde{\mathbf{v}} = 0$, Eq.8 can be written as Eq.9:

$$E = \frac{\rho}{2}\int \tilde{v}^2 dV + \frac{\rho}{2}\int \nabla.\left((\varphi + \tilde{\mathbf{v}}.\mathbf{r})(\tilde{\mathbf{u}} - \tilde{\mathbf{v}})\right)dV \qquad (9)$$

The first integral is equal to $\frac{\rho}{2}\tilde{v}^2(\pi R^2 - \pi R_0^2)$. By using the Green-Ostrogradsky theorem, the second integral can be replaced by the integral of $(\varphi + \tilde{\mathbf{v}}.\mathbf{r})(\tilde{\mathbf{u}} - \tilde{\mathbf{v}}).\mathbf{n}$ over the surfaces of the cylinders of radius $R$ and $R_0$. From the continuity equations, it is known that at the surface of the cylinder, $\tilde{\mathbf{v}}.\mathbf{n} = \tilde{\mathbf{u}}.\mathbf{n}$, and that the integration over the cylinder of radius $R_0$ is null. After substituting $\tilde{\mathbf{u}}$ and $\varphi$ by their expression (5) and (6) and neglecting the terms which approach zero when $R \to \infty$, Eq.9 now reads:

$$E = \frac{\rho}{2}\pi R_0^2 \tilde{v}^2 \qquad (10)$$

By application of the kinetic power theorem to Eq.10, a force $\mathbf{F}$ called the *trail force* on the cylinder can be calculated:

$$E = -\mathbf{F}.\tilde{\mathbf{v}} = -\left(\rho\pi R_0^2 \frac{d\tilde{\mathbf{v}}}{dt}\right).\tilde{\mathbf{v}} \qquad (11)$$

In the laboratory referential, we can replace $\tilde{\mathbf{v}}$ by $\mathbf{v} - \mathbf{u_0}$ with $\mathbf{v}$ velocity of the cylinder and $\mathbf{u_0}$ velocity of the fluid far from the cylinder. $\mathbf{F}$ is then equal to:

$$\mathbf{F} = \rho\pi R_0^2\left(\frac{d\mathbf{u_0}}{dt} - \frac{d\mathbf{v}}{dt}\right) \qquad (12)$$

Newton's second law states that the acceleration of the cylinder is equal to the sum of the trail force and a force related to the displacement of a volume of fluid equal to the one occupied by the cylinder:

$$\rho_0 \pi R_0^2 \frac{d\mathbf{v}}{dt} = \rho\pi R_0^2\left(\frac{d\mathbf{u_0}}{dt} - \frac{d\mathbf{v}}{dt}\right) + \rho\pi R_0^2 \frac{d\mathbf{u_0}}{dt}$$
$$(\rho_0 + \rho)\frac{d\mathbf{v}}{dt} = 2\rho \frac{d\mathbf{u_0}}{dt} \qquad (13)$$

After integration over $t$, we find the relationship 14 between $\mathbf{v}$ and $\mathbf{u_0}$:

$$\mathbf{v} = \frac{2\rho}{\rho_0 + \rho}\mathbf{u_0} \qquad (14)$$

Note that if the cylinder density is equal to the fluid density, both the cylinder and the fluid move at the same velocity.

Under the assumptions detailed in the last section, the measured signal generated by the Lorentz force is then proportional to the incident ultrasound wave, and makes the use of the technique described in this paper appropriate for observing ultrasound fields. However, the measure is performed all along the exposed section of the wire, as opposed to a single point with a classic hydrophone, and additional modeling is required to confer spatial resolution to this technique.

The Fourier transform of Eq.1 leads to Eq.15 where $\delta$ is the coordinate along a direction $k$ of angle $\theta$ with the X axis, $k$ being defined as $k = x.cos\theta + y.sin\theta$. Note that the direction of the magnetic field is always assumed always perpendicular to the wire, thus in the direction perpendicular to $k$.

$$E(k,\theta) = \int v_z B_\delta e^{-j2\pi\delta k} d\delta \qquad (15)$$

The velocity at each point is retrieved by taking the 2D inverse Fourier transform of Eq.15:

$$v(x,y) = \frac{1}{B}\int_0^\pi \int_{-\infty}^{+\infty} E(k,\theta)|k|e^{j2\pi\delta k} d\theta dk \qquad (16)$$

The measurement of the intensity along a direction $k$ for different angles $\theta$ allows thus the reconstruction of the velocity field picture.

The electromagnetic hydrophone measures thus primarily the ultrasound velocity, as given by Eq.16. This hydrophone could besides give the accurate intensity of ultrasound if combined with a pressure hydrophone[10].

For most ultrasound field measurements, the plane wave hypothesis is accurate, especially in the vicinity of the focus of convergent beams. In this case, peak pressure and velocity of the medium are linked, as previously described in Eq.2, and Eq.16 leads to:

$$p(x,y) = \frac{\rho c}{B} \int_0^\pi \int_{-\infty}^{+\infty} E(k,\theta)|k|e^{j2\pi\delta k} \, d\theta \, dk \qquad (17)$$

The setup diagram is presented in Figure 1. Experiments were conducted in a 48x48x20 cm$^3$ tank filled with degassed water at a temperature of 17±2 °C. A focused transducer is used to create an ultrasound wave propagating in the Z axis. The magnetic field placed is along the X axis, perpendicular to the ultrasound wave. A wire is placed along the Y axis, perpendicular to the ultrasound wave and the magnetic field.

A generator (Agilent HP33120A, Santa Clara, CA, USA) creates a 10 V peak-to-peak signal at 1.1 MHz with 2 sinusoids per pulse at a pulse repetition frequency of 100 Hz. The signal is amplified by a 200W amplifier (0.5-100MHz, Kalmus Engineering LA200H model, Rock Hill, SC, USA). A 1.1 MHz, 50 mm in diameter, focused at 50 mm, air-backed transducer is converting the electric signal into an ultrasound wave. The transducer displacement is driven by a PC-based motor (Newport Corporation MM4005, Irvine, CA, USA).

The sensitive part of the hydrophone is an insulated copper wire placed at focal distance (50 mm) from the transducer. It is brazed to a BNC cable on both ends and is loosely held. The wire diameter is 217±7 μm with a length of 12 cm. The resistance between the two extremities is 0.8±0.2 Ohm at 1.1 MHz. The signal is amplified by a 10$^6$ V.A$^{-1}$ current amplifier (Laser Components HCA-2M-1M, Olching, Germany) with an input impedance of 50 Ω, then filtered from 500 kHz to 2 MHz with a bandpass filter (NF Corporation FV-628B, Yokohama, Japan), and is observed with a PC-based digital oscilloscope (LeCroy WaveSurfer 422, Chestnut Ridge, NY, USA).

The magnetic field is created by a U-shaped permanent magnet. Each pole is composed by two 5x5x3 cm$^3$ NdFeB magnets (BLS Magnet, Villers la Montagne, France). The gap between the poles is 4.5 cm. The magnetic field is 200±20 mT along a length of 2 cm around the middle of the wire.

The translations and rotations were performed by the transducer instead of the hydrophone for reasons of maneuverability. The transducer was moved in rotation from 0 to 179° with a 1° step around its center and in translation on each angle from -4 to 4 mm with a 0.1 mm step. The peak-to-peak amplitude was acquired for each rotation and translation. A sinogram (position vs. angle picture) is created from each experiment. The picture is reconstructed through inverse Radon transform using the *iradon* Matlab function (The MathWorks, Natick, MA, USA), combined with a linear interpolation and a Hann filter for decreasing high frequency noise.

Tomography results are compared with standard piezoelectric needle hydrophone measurements (Speciality Engineering Associates, model reference: PZT-Z44-0400-H816, Sunnyvale, CA, USA) with a 400 μm diameter sensitive surface and a sensitivity of 1.5 10$^{-8}$ V.Pa$^{-1}$ at 1.1 MHz. It is placed in a similar position to the electromagnetic hydrophone at 50 mm from the transducer and moved perpendicularly to the ultrasound axis from -4 to 4 mm with a step of 0.1 mm on X and Y axis, roughly centered on the focal spot. The signals were acquired by a PC-based oscilloscope (LeCroy WaveSurfer 422, Chestnut Ridge, NY, USA) for 2D field reconstruction.

Alternately, the waveforms acquired by both hydrophones have been recorded in the same conditions except the use of 15 sinusoids per pulse and the absence of bandwidth filter.

The contour plot of the acoustic pressure obtained by electromagnetic hydrophone with 180 rotations and by piezoelectric hydrophone are drawn on Figure 2. Waveforms with normalized amplitude from the piezoelectric hydrophone and from the electromagnetic hydrophone are represented on Figure 3. The piezoelectric hydrophone gives an amplitude pressure at the focal point equal to 1.3 MPa. The maximum voltage found on the reconstructed picture of the electromagnetic hydrophone is equal to 62±2 mV. The focal spot has a width (measured at -3 dB of the peak pressure) of 1.23 mm for the piezoelectric hydrophone and 1.57 mm for the electromagnetic hydrophone, as shown on Figure 4. Each acquisition takes approximately 2 s. With 180 rotations and 80 translations per angle, the whole tomographic acquisition runs for over 8 hours. The tomographic reconstruction computing duration (< 1 s) is almost negligible compared to the acquisition time. The total acquisition time for the piezoelectric hydrophone, with 80x80 translations and 2 s per acquisition is about 4 hours.

If we consider the magnetic field as constant over a pixel, the pressure is given by Eq.18:

$$p = \frac{\rho c}{B} \frac{eR}{l\alpha} K \quad (18)$$

with the density of water $\rho = 1000\pm10$ kg.m$^{-3}$, the speed of sound in water $c = 1470\pm10$ m.s$^{-1}$, the measured voltage $e = 0.062\pm0.002$ V at the focal point, resistance of the wire plus amplifier $R = 50.8\pm0.2$ Ω, the magnetic field on the wire $B = 0.20\pm0.02$ T, the width of 1 pixel $l = 0.1$ mm the amplification of signal $\alpha = 10^6$ V.A$^{-1}$, the correction factor given by Eq.14 $K = (\rho+\rho_0)/2\rho = 5.0\pm0.3$ with $\rho_0 = 9000\pm500$ kg.m$^{-3}$ (density of insulatedcopper ), the corresponding peak-to-peak pressureis equal to 1.2±0.3 MPa. This result is in very good agreement with the measurement of the piezoelectric hydrophone without ultrasound calibration need.This result confirms our hypotheses described in the section devoted to signal measurement theory and demonstrates the feasibility of making quantitative measurements with this technique. The strength of the theoretical model that gives the correction factor is to be mainly based on one major hypothesis: that diameter of the wire is sufficiently small compared to the ultrasound wavelength.

The 0.5-2 MHz filter limits our frequency bandwidth, but doesn't change the amplitude of the acquired result. The waveform acquired by the piezoelectric hydrophone and the electromagnetic hydrophone are nevertheless almost identical at 1.1 MHz. The theoretical study presents a relationship between the fluid velocity and the wire velocity without transition time. This relationship indicates that the wire should follow the fluid movement well, at least when the conditions proposed here are fulfilled.

The electromagnetic hydrophone creates signal all along the parts exposed to ultrasound, increasing its sensitivity compared to the previous designs of electromagnetic hydrophones.The number of acquisitions is given by the number of translations $n_t$ times the number of rotations $n_r$ giving the quality of the reconstruction. We find experimentally that $n_r = 180$ is a good compromise between quality of reconstruction and speed of acquisition. It has to be compared to the number of acquisitions needed by a standard piezoelectric hydrophone, equal to $n_t^2$ for a square picture. The spatial resolution of the electromagnetic hydrophone is mainly given by the size of the step displacement. A difference of 20% in the focal spot size has been observed between the two techniques. As the variation of pressure at millimeter scale is seen, the spatial resolution has beenconsidered assubmillimetric. However, additional measurements are required to obtain a quantitative estimation.

Without a Faraday cage, the hydrophone can still be sensitive to electromagnetic interference.

The electromagnetic hydrophone is fairly inexpensive, the main cost coming from the magnet which is not necessary if a magnetic field is present (in MRI for example).

The authors would like to thank Amalric Montalibet for his work on acousto-magnetic interaction, Adrien Matias for the mechanical fabrication, Alain Birer for the electronic help, Frédéric Padilla, Rémi Souchon, Stefan Catheline for their advices and Andrew Fowler for its help with the language.

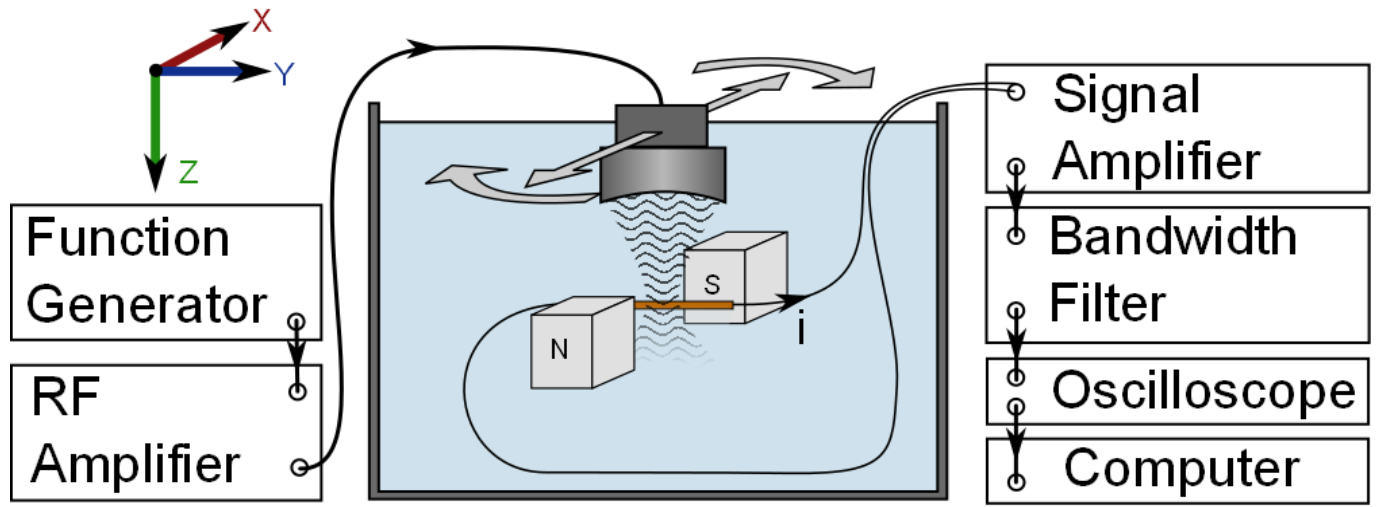

Figure 1 Scheme of the experiment. *For clarity reason the poles of the magnet have not been represented.*

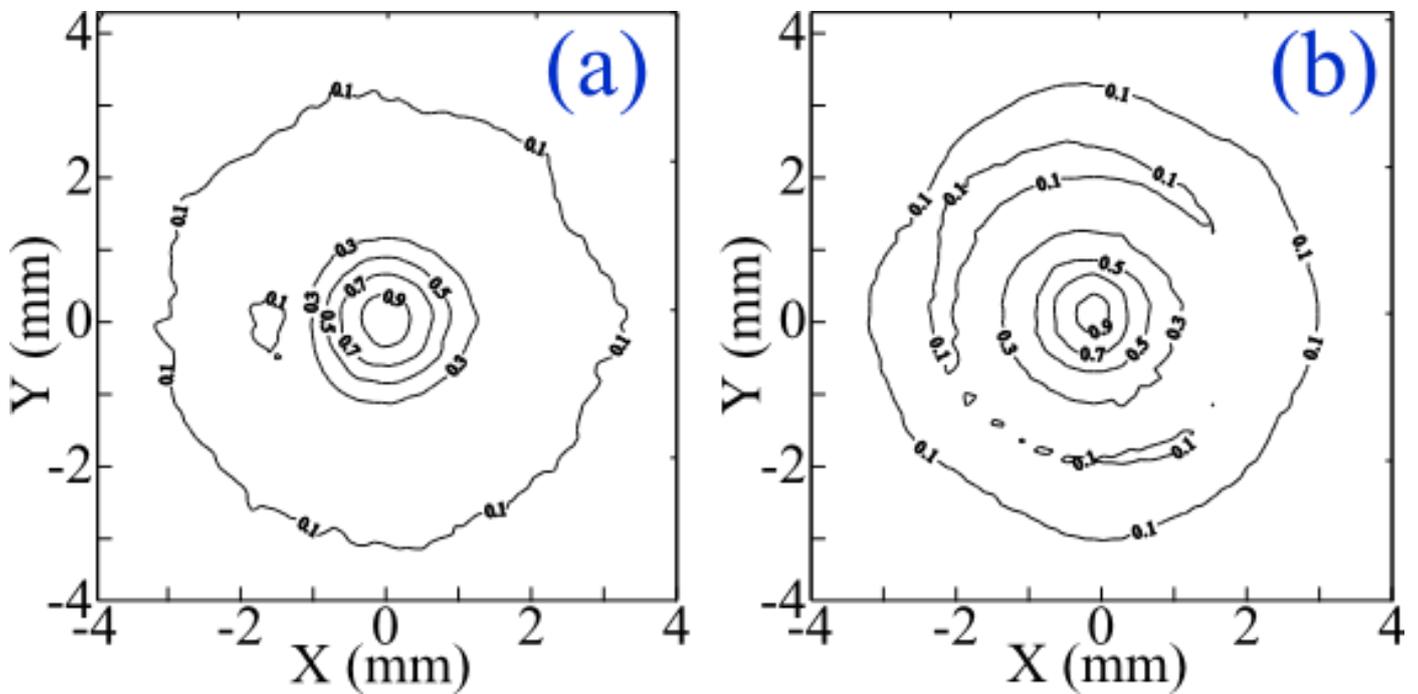

Figure 2 (a) Contour of pressure field acquired by electromagnetic hydrophone reconstructed with 180 rotations of the wire (b) Contour of pressure field acquired by piezoelectric hydrophone

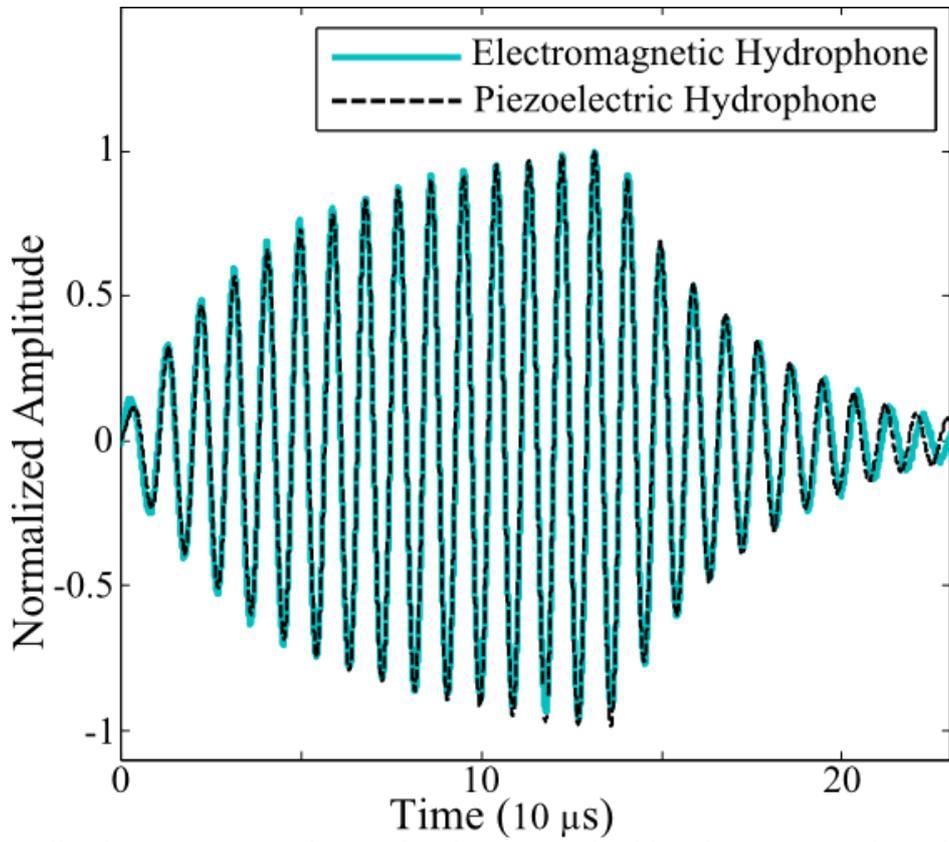

Figure 3 Normalized pressure over time at focal spot acquired by electromagnetic and piezoelectric hydrophone

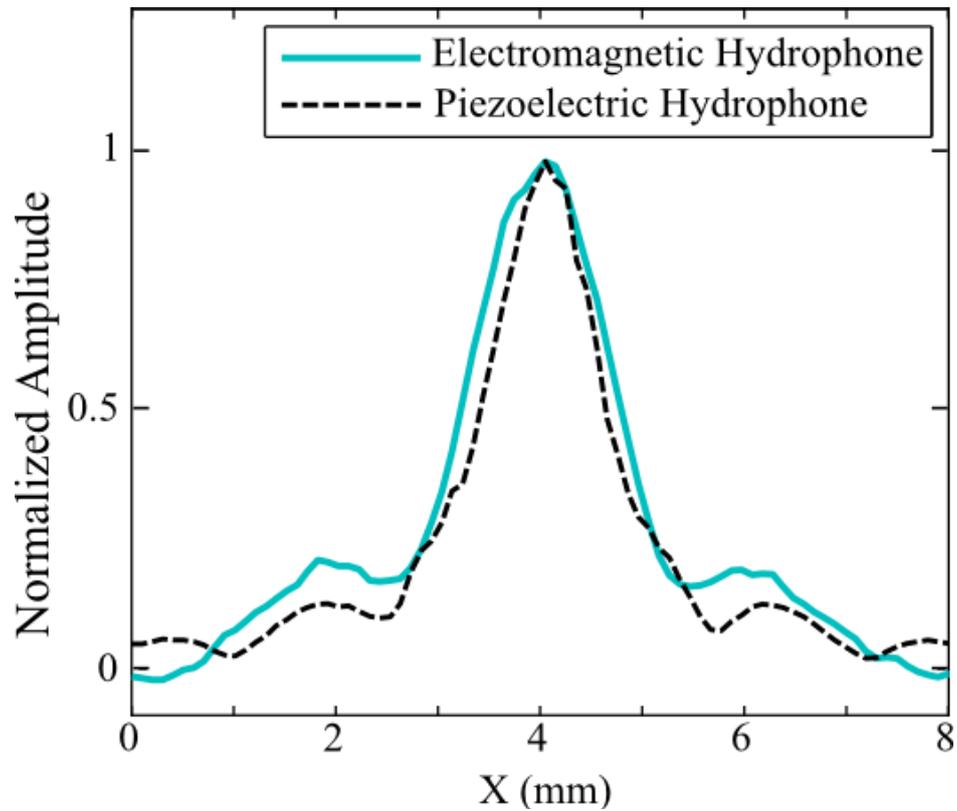

Figure 4 Normalized pressure along a line through focal spot acquired by piezoelectric and electromagnetic hydrophone